\documentclass[aps,prl,superscriptaddress,twocolumn,floatfix,showpacs,amsfonts,amssymb,amsmath]{revtex4}
\usepackage{epsfig} 
\usepackage{amssymb} 
\usepackage{amsmath} 
\usepackage{amsfonts} 
\usepackage{color, graphicx} 
\usepackage{bm}

\def\be{\begin{equation}} 
\def\ee{\end{equation}} 

\def\Tr{{\rm Tr}}

\begin{document}
\title{Circumventing the odd particle-number sign problem in the shell model Monte Carlo}

\author{Y.~Alhassid}
\affiliation{Center for Theoretical Physics, Sloane Physics Laboratory,Yale University, New Haven, CT 06520, USA}
\author{P.~Fanto}
\affiliation{Center for Theoretical Physics, Sloane Physics Laboratory,Yale University, New Haven, CT 06520, USA}
\author{C.~\"{O}zen}
\affiliation{Faculty of Engineering and Natural Sciences, Kadir Has University, Istanbul 34083, Turkey}
\date{\today}
\begin{abstract}
The shell model Monte Carlo (SMMC) method is a powerful method for calculating exactly (up to statistical errors) thermal observables and statistical properties of atomic nuclei.  However, its application has been limited by a sign problem at low temperatures that arises from the projection onto odd particle number even for good-sign interactions.  Here, we develop a technique -- the partition function extrapolation method (PFEM) -- to extract the ground-state energy of an odd-mass nucleus from the excitation partition function calculated at temperatures at which this sign problem is moderate.  We validate the PFEM in heavy even-mass nuclei and systematically calculate ground-state energies for isotopic chains of heavy odd-mass nuclei.  The PFEM can be extended to other finite-size quantum many-body systems. 
\end{abstract}


\maketitle

{\it Introduction} -- The shell model Monte Carlo (SMMC) method~\cite{johnson1992, lang1993, alhassid1994, koonin1997,alhassid2008, alhassid_rev} can calculate exactly (up to statistical errors) finite-temperature observables within the configuration-interaction (CI) shell model framework in model spaces that are far beyond the reach of conventional diagonalization methods.  The SMMC has been applied to study nuclear state and level densities in nuclei as heavy as the lanthanides~\cite{ozen2013, bonett2013, alhassid2014, ozen2015, alhassid2015, mustonen2018}.  Similar auxiliary-field Monte Carlo (AFMC) methods have been applied to other quantum many-body systems such as cold atoms~\cite{magierski2009,gilbreth2013,jensen2020}.

Quantum Monte Carlo calculations for fermions are often limited by the Monte Carlo sign problem, which leads to large statistical errors.  Calculations of state and level densities~\cite{ozen2013, bonett2013, alhassid2014, ozen2015, alhassid2015, mustonen2018} have been carried out using interactions that include the dominant pairing and multipole-multipole components of effective nuclear interactions~\cite{dufour1996}, which have no sign problem in the grand-canonical ensemble.  However, the projection onto fixed numbers of protons and neutrons is necessary for describing finite-size nuclei. While the projection on an even number of particles (both protons and neutrons) preserves the good sign of the interaction, the projection onto odd particle number introduces a sign problem at low temperatures, which we shall refer to as the odd-particle sign problem. 

Although this odd-particle sign problem is moderate at not too low temperatures, the statistical fluctuations become too large at low temperatures for the ground-state energies in odd particle-number systems to be directly determined.  The ground-state energy is a crucial quantity because it is necessary for determining the excitation energy in level density calculations.  In Ref.~\cite{mukherjee2012},  imaginary-time Green's functions of neighboring even-number systems were used to determine ground-state energies of odd-mass nuclei.  However, calculating these Green's functions in heavy nuclei is computationally expensive.  In Ref.~\cite{ozen2015}, SMMC calculations at higher temperatures were combined with experimental results to extract the ground-state energy of odd-mass nuclei.  

Here, we introduce a self-contained method -- the partition function extrapolation method (PFEM) -- that determines the ground-state energy from the SMMC partition function at temperatures above the onset of the sign problem.  This method consists of expressing the excitation partition function via a parameterized model for the state density and fitting the parameters of this model to obtain the ground-state energy.  We validate this method by applying it to even-even nuclei, in which the ground-state energy can be determined directly from SMMC calculations at low temperatures.  We then apply the PFEM systematically to calculate the ground-state energies of odd-mass samarium and neodymium isotopes.  

Beyond the SMMC, the PFEM is also useful for determining ground-state energies in the static-path plus random-phase approximation (SPA+RPA), which has recently been applied to calculate state densities of heavy nuclei~\cite{fanto2021}.  The SPA+RPA breaks down at low but nonzero temperature, requiring the use of an extrapolation method to determine the ground-state energy.  In Ref.~\cite{fanto2021}, a preliminary version of the PFEM was applied to extract the SPA+RPA ground-state energies of samarium isotopes $^{148-155}$Sm.

{\it Method} -- In the SMMC method, the thermal expectation value of an observable $\hat O$ is given by \cite{alhassid_rev}
\be\label{smmc}
\langle \hat O \rangle = \frac{\Tr \left( \hat O e^{-\beta \hat H}\right)}{\Tr\left( e^{-\beta \hat H}\right)} = \frac{\int D[\sigma] W_\sigma \Phi_\sigma \langle \hat O \rangle_\sigma}{\int D[\sigma] W_\sigma \Phi_\sigma}\,,
\ee
where $\hat H$ is the nuclear Hamiltonian, $\sigma$ are a set of time-dependent auxiliary fields, $W_\sigma = G_\sigma | \Tr_{\mathcal{A}} \hat U_\sigma|$ is a positive-definite weight function ($G_\sigma$ is a Gaussian weight and $\hat U_\sigma$ is the propagator of non-interacting nucleons moving in external auxiliary fields $\sigma$), $\Phi_\sigma = \Tr_{\mathcal{A}} \hat U_\sigma/ | \Tr_{\mathcal{A}} \hat U_\sigma|$ is the Monte Carlo sign, and $\langle \hat O \rangle_\sigma = \Tr_{\mathcal{A}}\,(\hat O \hat U_\sigma)/\Tr_{\mathcal{A}} \hat U_\sigma$. Here  $\Tr_{\mathcal{A}}$ represents the canonical-ensemble trace with respect to fixed numbers of protons and neutrons and is calculated with projection methods; see Ref.~\cite{alhassid_rev} and works cited therein. 
In the SMMC method, the Metropolis-Hastings algorithm is used to sample uncorrelated field configurations $\sigma_k$, and observables are estimated by averages over the samples, i.e., $\langle \hat O \rangle \approx \sum_k \Phi_{\sigma_k} \langle \hat O \rangle_{\sigma_k}/\sum_k \Phi_{\sigma_k}$.  For odd number of protons or neutrons, the average sign $\langle \Phi_\sigma \rangle$  decays rapidly as the temperature is lowered, leading to large statistical errors on the Monte Carlo estimates of observables at low temperatures.

The goal of the PFEM is to determine the ground-state energy $E_0$, given the SMMC thermal energy estimates and their associated errors at higher temperatures.  The thermal energy $E(\beta)$ at inverse temperature $\beta$ can be calculated in the SMMC as the expectation value of the Hamiltonian $\hat H$.  Measuring the energies relative to a reference energy $E_{\rm ref}$,  the corresponding excitation partition function is given by
\be\label{Z_ex}
Z^\prime(\beta ; E_{\rm ref}) = Z(\beta) e^{\beta E_{\rm ref}}\,,
\ee
where the partition function $Z(\beta)={\rm Tr}\, e^{-\beta \hat H}$ can be obtained by the integral relation $\ln Z(\beta) = \ln Z(0) - \int_0^\beta d\beta^\prime E(\beta^\prime)$ ($\ln Z(0)$ depends only on the single-particle model space dimension and numbers of valence nucleons).  The excitation partition function for an arbitrary reference energy $E_{\rm ref}$  can be related to the excitation partition function for the ground state energy $E_0$ by
\be\label{Z_ex_relation}
\ln Z^\prime(\beta ; E_{\rm ref}) = \ln Z^\prime(\beta; E_0) - \beta(E_0 - E_{\rm ref})\,.
\ee

We express $Z^\prime(\beta; E_0)$ in Eq.~(\ref{Z_ex_relation}) as the Laplace transform of the state density $\rho(E_x)$
\be\label{Z_0}
Z^\prime(\beta; E_0) = \int_0^\infty dE_x \,\rho(E_x) e^{-\beta E_x}\,,
\ee
where $E_x = E - E_0$ is the excitation energy.  In the PFEM, we use a parameterized model for $\rho(E_x)$ in Eq.~(\ref{Z_ex_relation}) and fit the parameters of the model, together with the ground state energy $E_0$, to the SMMC results for $\ln Z^\prime(\beta ; E_{\rm ref})$.  The SMMC results can be calculated for not too low temperatures for which the sign problem is moderate.  In particular, to describe odd-mass nuclei, we use the back-shifted Bethe formula (BBF)~\cite{dilg1973}
\be\label{bbf}
\rho_{\rm BBF}(E_x) = \frac{\sqrt{\pi}}{12 a^{1/4}} \frac{e^{2\sqrt{a(E_x-\Delta)}}}{(E_x-\Delta)^{5/4}}\;,
\ee
where $a$ is the single-particle level density parameter and $\Delta$ is the back-shift parameter.
Inserting Eq.~(\ref{bbf}) into Eq.~(\ref{Z_0}) and then inserting that result into Eq.~(\ref{Z_ex_relation}) expresses the SMMC excitation partition function in terms of the three parameters $(a,\Delta,E_0)$.  
We can then determine these parameters by a $\chi^2$ fit.  

In practice, we carry out this fit in two steps.  In the first step, we apply the saddle-point approximation to the integral in Eq.~(\ref{Z_0}), in which we use  for $\rho(E_x)$ the BBF in Eq.~(\ref{bbf}), and obtain
\be\label{saddle_point}
\ln Z^\prime(\beta; E_{\rm ref}) \approx \frac{a}{\beta} + \ln\left(\frac{\pi\beta}{6a}\right) - \beta S\,,
\ee
where $S = E_0 - E_{\rm ref} + \Delta$.  We fit Eq.~(\ref{saddle_point}) to the SMMC data at moderate temperatures to obtain fitted values of $(a,S)$.  In the second step, keeping the values of $(a,S)$ fixed, we fit the full expression (\ref{Z_ex_relation}) to the SMMC data at low temperatures, where we again use the BBF in Eq.~(\ref{bbf}) for $\rho(E_x)$ and carry out the integral in Eq.~(\ref{Z_0}) numerically.  This is a one-parameter $\chi^2$ fit of $E_0$; the back-shift parameter $\Delta$ is determined by $\Delta = S - (E_0 - E_{\rm ref})$.

For positive values of $\Delta$, the BBF (\ref{bbf}) is not defined for $E_x\leq \Delta$. 
In such cases, we replace the BBF in Eq.~(\ref{Z_0}) with the Gilbert-Cameron composite formula \cite{gilbert1965}
\be\label{composite}
\rho_{\rm comp}(E_x) = \begin{cases} \frac{1}{T_1} e^{(E_x - E_1)/T_1} \;\;\; & E_x < E_M \\
\rho_{\rm BBF}(E_x)  & E_x > E_M
\end{cases}\,,
\ee
where $E_M$ is a matching energy and $(E_1,T_1)$ are defined by the conditions that the state density and its first derivative be continuous at $E_M$.  In this case, the second step of the PFEM is a two-parameter fit of the values $(E_0, E_M)$.  For the odd-mass isotopes studied here, we find that $\Delta$ is negative in the BBF.  However, the composite formula was applied to the SPA+RPA in Ref.~\cite{fanto2021}.  

{\it Validation} -- To assess the accuracy of the PFEM, we applied it to calculate the ground-state energies of even-mass samarium isotopes $^{148,150,152,154}$Sm.  For these nuclei, there is no sign problem and we can calculate $E_0$ from the thermal SMMC energies at large $\beta$ values.  We used the model space and effective pairing plus multipole-multipole interaction described in Ref.~\cite{ozen2013}.  For $^{152,154}$Sm, the back-shift $\Delta$ is negative, as is typical for odd-mass nuclei, and we use the BBF (\ref{bbf}) in the second step of the PFEM.  For $^{148,150}$Sm, $\Delta$ is positive and we use instead the composite formula (\ref{composite}) in the second step.

\begin{figure}[ht!]
\includegraphics[width=0.5\textwidth]{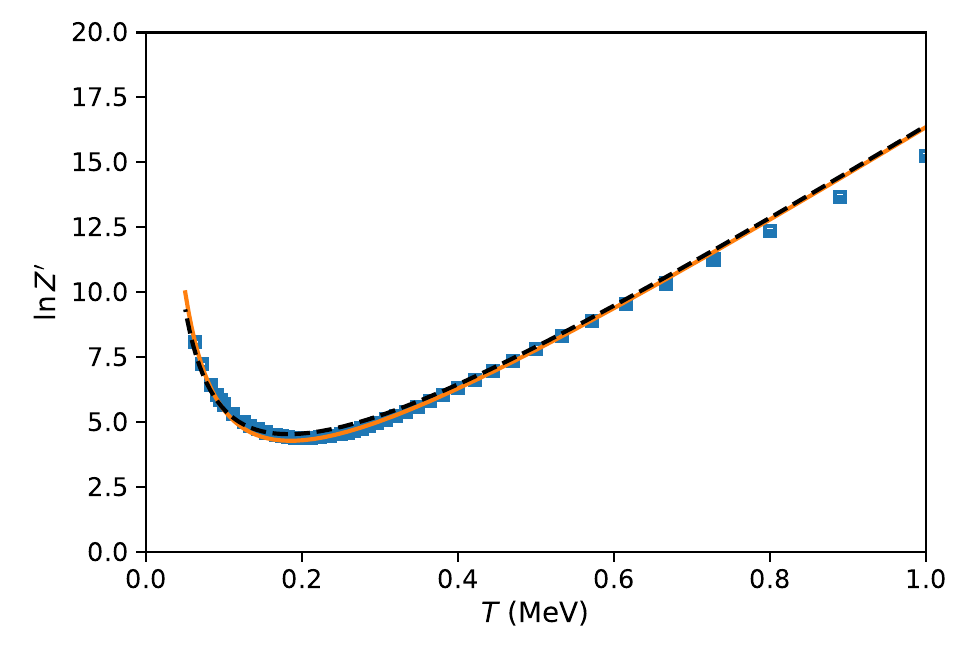}
\caption{\label{sm154_lnZprime_fig} The logarithm of the excitation partition function $\ln Z^\prime$ as a function of temperature $T$ for $^{154}$Sm.  The SMMC results (blue squares) are compared with the saddle point fit results (orange line) and BBF fit results (black dashed line).}
\end{figure}

\begin{table*}
\centering
\caption{\label{even_table} The ground-state energies $E_0^{\rm PFEM}$ from the PFEM compared with the ground-state energies $E_0^{\rm SMMC}$ obtained from SMMC energies at high $\beta$ values.  For $^{152,154}$Sm, we obtain $E_0^{\rm SMMC}$ by a fit to a rotor model, while for $^{148,150}$Sm we average the energies for $\beta \approx 8-15$ MeV$^{-1}$.  In the PFEM, we use the BBF in the second step for $^{152,154}$Sm and the composite formula in the second step for $^{148,150}$Sm.  The value $E_0^{\rm PFEM}$ is obtained using $\beta \approx 2.5 - 15$ MeV$^{-1}$ in the second step of the fit, while $E_0^{\rm PFEM, \, r}$ is obtained with $\beta$ restricted to the range $\beta \approx 3.5-5$ MeV$^{-1}$ typical of an odd-mass nucleus.}
\begin{tabular}{l c c c}
\hline\hline
 & $E_0^{\rm SMMC}$ (MeV)  & $E_0^{\rm PFEM}$ (MeV) & $E_0^{\rm PFEM, \, r}$ (MeV) \\\hline 
 $^{154}$Sm & -295.45 $\pm$ .01 & -295.43 ($+.08$,$-.03$)  & -295.21 ($+.11$, $-.11$) \\
 $^{152}$Sm & -275.85 $\pm$ .02 & -275.77 ($+.05$,$-.02$) & -275.55 ($+.10$,$-.11$) \\
 $^{150}$Sm & -255.77 $\pm$ .02 & -255.99 $\pm$ .14 & -256.08 $\pm$ .47  \\
$^{148}$Sm & -235.66 $\pm$ .02 & -235.95 $\pm$ .27 & -235.92 $\pm$ .40 \\ 
\hline\hline
\end{tabular}
\end{table*}

As a typical result, we show in Fig.~\ref{sm154_lnZprime_fig} the logarithm of the excitation partition function (\ref{Z_ex}) of $^{154}$Sm calculated in the SMMC, compared with the fits from the saddle-point formula (\ref{saddle_point}) and the full BBF (\ref{bbf}) used in Eq.~(\ref{Z_0}).  The saddle-point formula is fitted to data in the moderate temperature range $0.3 \le T \le 0.6$ MeV, while the full BBF is fit to low-temperature data $T \le 0.4$ MeV.  

In Table ~\ref{even_table}, we compare the PFEM ground-state energies with those extracted from SMMC energies at high $\beta$ values.  We estimated the errors on the PFEM ground-state energies differently for the BBF and composite formula approaches.  For the BBF, the error bar on $E_0$ reflects the points at which the $\chi^2$ per degree of freedom changes by one from its minimum value obtained at the fitted $E_0$ point.  This change is asymmetric, and this asymmetry reflects the constraint that $\Delta$ be negative.  For the composite formula approach applied to $^{148,150}$Sm, we instead calculated the $\chi^2$ per degree of freedom for multiple $E_M$ values and fit a parabola to these results.  We then calculate the error bars on $E_M$ and $E_0$ by finding the points at which this parabola predicts the $\chi^2$ per degree of freedom to increase by one.  This latter approach results in significantly higher error bars, limiting the composite fit approach.  For practical applications, the limitation is not significant as the BBF can be applied to nearly all odd-mass isotopes.

For the more deformed isotopes $^{152,154}$Sm, we calculated the reference SMMC ground-state energies $E_0$ in Table~\ref{even_table} by fitting the thermal SMMC energies at high $\beta$ to a rotational model $E(T) \approx E_0 + T$ \cite{ozen2013}.  In contrast, for $^{148,150}$Sm, we determined $E_0$ by averaging the SMMC energies in the range $\beta \approx 8-15$ MeV$^{-1}$.  We find excellent agreement between the PFEM results for $^{152,154}$Sm, for which we used the BBF in the second step of the fit, and the reference SMMC values.  For $^{148,150}$Sm, for which we used the composite formula in the second step of the fit, we find that the PFEM ground-state energies are systematically lower than the reference SMMC values by roughly $\sim 300$ keV. 

Finally, we note that, for odd-mass nuclei, we cannot calculate reliably the SMMC excitation partition function at low temperatures, and it is useful to examine how this temperature restriction affects the PFEM results.   
Table \ref{even_table} shows the values $E_0^{\rm PFEM,\,r}$ obtained for the ground-state energies using the restricted temperature range $\beta \approx 3.5-5$ MeV$^{-1}$ in the second step of the fit.  Restricting the temperature range shifts the BBF fit to somewhat higher values, as shown in the results for $^{152,154}$Sm.  In contrast, the temperature restriction does not affect much the results of the composite formula fit.  In both the BBF and composite formula cases, the temperature restriction increases the size of the error bars.  This result emphasizes the importance of extending the odd-mass SMMC calculations to as large $\beta$ values as possible. 

\begin{figure}[ht!]
\includegraphics[width=0.5\textwidth]{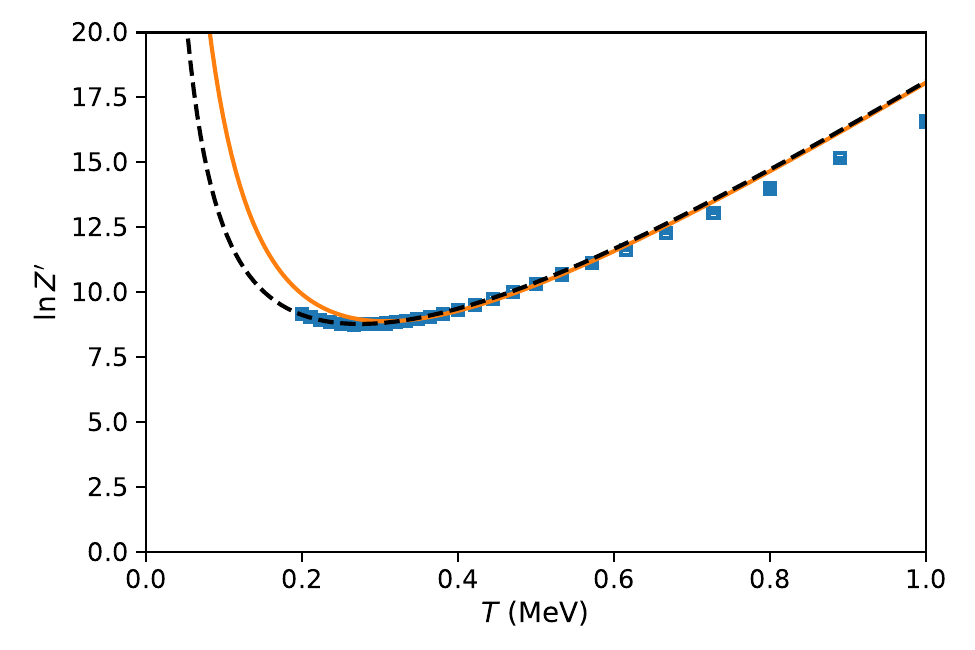}
\caption{\label{sm153_lnZprime_fig} The logarithm of the excitation partition function $\ln Z^\prime$ as a function of temperature $T$ for $^{153}$Sm.  Colors and symbols are as in Fig.~\ref{sm154_lnZprime_fig}.}
\end{figure}

\begin{table*}[ht!]
\caption{\label{odd_table} The ground-state energy $E_0^{\rm PFEM}$  determined by the PFEM for the odd-mass neodymium and samarium isotopes, compared with the  ground-state energy $E_0^{\rm exp}$ obtained in the method of Ref.~\cite{ozen2015}, which combines the SMMC calculations with experimental data.   The BBF is used in the second step of the PFEM in each case. We also show the parameters $a,\Delta$ extracted in the PFEM.}
\begin{tabular}{l c c c c c c}
\hline\hline
 & $E_0^{\rm PFEM}$ (MeV) & $E_0^{\rm exp}$ (MeV) & $a^{\rm PFEM}$ (MeV$^{-1}$) & $\Delta^{\rm PFEM}$ (MeV) \\ \hline
$^{143}$Nd & -191.71 ($+.06$,$-.03$) & -191.61 $\pm$ .02 & 12.97 $\pm$ .12 & -.06 ($+.03$,$-.06$) \\ 
$^{145}$Nd & -210.11($+.06$,$-.02$) &  -210.01 $\pm$ .03 & 16.42 $\pm$ .10 &  -.03 ($+.02$,$-.06$) \\
$^{147}$Nd & -228.00 ($+.21$,$-.05$) & -227.87 $\pm$ .03 & 18.46 $\pm$ .07 & -.06 ($+.05$,$-.21$)  \\
$^{149}$Nd & -245.48 ($+.12$,$-.13$) & -245.44 $\pm$ .02 & 19.94 $\pm$ .03 & -.37 ($+.13$,$-.12$)  \\
$^{151}$Nd & -262.92 ($+.11$,$-.12$) & -262.50 $\pm$ .03 & 20.09 $\pm$ .06 & -.51 ($+.12$,$-.11$)  \\
$^{149}$Sm & -244.83 ($+.18$,$-.15$) & -244.90 $\pm$ .03 & 18.87 $\pm$ .04 & -.19 ($+.15$, $-.18$) \\
$^{151}$Sm & -264.94 ($+.08$,$-.08$) & 264.97 $\pm$ .02 & 20.10 $\pm$ .06 &  -.45 ($+.08$,$-.08$)  \\
$^{153}$Sm & -284.89 ($+.13$,$-.14$) & -284.80 $\pm$ .02 & 20.13 $\pm$ .08 & -.69 ($+.10$,$-.10$) \\
$^{155}$Sm & -304.57 ($+.16$,$-.18$) & -304.25 $\pm$ .03 & 19.59 $\pm$ .10 & -.69 ($+.18$,$-.16$) \\
\hline\hline 
\end{tabular}
\end{table*}

{\it Application to odd-mass lanthanides} -- Having validated the PFEM for even-mass nuclei, we applied this method to calculate the ground-state energies of odd-mass neodymium and samarium isotopes.  We used the BBF in the second step of the fit in each case.  Fig.~\ref{sm153_lnZprime_fig} shows a representative result for $^{153}$Sm.  As shown in this figure, the sign problem limits the SMMC data (solid squares) to $\beta$ values below $\beta \approx 4-5$ MeV$^{-1}$.  The solid line describes the first step of fitting the saddle-point formula to the SSMC data in a moderate temperature range $0.3 \le T \le 0.6$ MeV to determine $a$ and $S$. The dashed line is the BBF fit in the low-temperature range $0.2 \le T \le 0.4$ MeV (where SMMC data exists) to determine the ground-state energy $E_0$.

In Table~\ref{odd_table}, we compare the PFEM results to those obtained from the method of Ref.~\cite{ozen2015}, which combines the SMMC results with experimental data to extract the average excitation energy as a function of temperature.  Overall, the ground-state energies of the two methods are in good agreement.  The PFEM $E_0$ values tend to be somewhat lower than those from the method of Ref.~\cite{ozen2015}.  Also, the PFEM gives larger error bars, especially for the more neutron-rich neodymium and samarium isotopes.  The main advantage of the PFEM is that it does not rely on any experimental data and uses only the SMMC calculations.

{\it Conclusions and Outlook} -- We have developed the partition function extrapolation method (PFEM) to extract the ground-state energy from SMMC calculations in odd-mass nuclei, which are limited at low temperatures by a sign problem resulting from the projection onto odd particle number.  The PFEM consists of applying a parameterized model for the state density to fit the excitation partition function (\ref{Z_ex}), which we obtain from SMMC calculations.  In the PFEM we primarily apply the BBF state density (\ref{bbf}) since the back-shift parameter $\Delta$ is usually negative for odd-mass nuclei, but in cases in which $\Delta$ is positive we use instead the composite formula (\ref{composite}).

We validated the PFEM method in even-even samarium isotopes, for which ground-state energies can be obtained directly from the SMMC thermal energies at low temperatures.
We then applied the PFEM to calculate the ground-state energies for odd-mass neodymium and samarium isotopes and found excellent agreement with the ground-state energies obtained with the method of Ref.~\cite{ozen2015}, which combined the SMMC results with experimental data.  A main advantage of the PFEM is that it requires no additional information beyond the SMMC results.  Moreover, the PFEM is computationally efficient, in contrast to the Green's function method of Ref.~\cite{mukherjee2012}.

The PFEM is also useful in the context of many-body methods other than the SMMC. For example, in Ref.~\cite{fanto2021}, a preliminary version of the PFEM was used to estimate ground-state energies in the static-path plus random-phase approximation (SPA+RPA).  

Finally, the PFEM can be extended to AFMC studies in many-body systems other than nuclei, provided that a reliable parameterized model for the state density can be found.  
A possible application would be to determine the energy staggering pairing gap in strongly interacting cold atomic  two-species Fermi gases~\cite{gilbreth2013,jensen2020}.  This gap is defined for $N$ particles by $\Delta_E  =[2E(N/2+1,N/2)- E(N/2+1,N/2+1) -E(N/2,N/2)]/2$, where $E(N_\uparrow, N_\downarrow)$  is the ground-state energy for $N_\uparrow$  spin-up and $N_\downarrow$ spin-down particles. For $N_\uparrow=N_\downarrow$ there is no sign problem but the spin-imbalanced system with  $N_\uparrow \neq N_\downarrow$ has a sign problem at low temperatures. 

{\it Acknowledgments} -- 
This work was supported in part by the U.S. DOE grant No.~DE-SC0019521, 
and by the U.S. DOE NNSA Stewardship Science Graduate Fellowship under cooperative agreement No.~NA-0003960.
The calculations used resources of the National Energy Research Scientific Computing Center (NERSC), a U.S. Department of Energy Office of Science User Facility operated under Contract No.~DE-AC02-05CH11231.  We thank the Yale Center for Research Computing for guidance and use of the research computing infrastructure.

\end{document}